# The role of diamagnetism in the separation of particles and sharp edges of the Saturn's ring

Vladimir V. Tchernyi[1], Sergey V. Kapranov[2]

[1]Modern Science Institute, SAIBR, Moscow, Russia, chernyv@bk.ru
[2]A.O. Kovalevsky Institute of Biology of the Southern Seas, RAS, Moscow, Russia, sergey.v.kapranov@yandex.ru

**Abstract**

The diamagnetism of ice particles of the rings can explain their separation and the sharp edges of the rings. The existing gravitational theories of the origin of rings can't explain these observed facts. Taking into account the magnetic field of Saturn, all the particles of the rings acquire stability in the horizontal and vertical directions. The force of diamagnetic expulsion of inhomogeneous magnetic field inside the rings structure forms sharp edges and separates the particles.

## 1. Introduction

We suggest explanations of sharp edges of Saturn's rings and separation of their particles. Earlier explanation of these phenomena in the rings structure is based on the synchronization phenomenon due to which the epicycle rotational phases of particles in the ring, under certain conditions, become synchronized with the phase of external satellite [1]. In accordance data of Cassini probe existing explanation is valid only for irregularity of the edges of rings due to gravitational interaction of particles with satellite. At the same time sharp edges of rings and separation of particles contains even under influence of fluctuation of the gravitational field of Saturn and creation of wave of density and surface waves in the rings structure.

Questions about many observed features of Saturn's rings have remained unsolved since G. Galilei first saw them in 1610. In 1856 J. Maxwell proved that rings are made of particles. In 1947 G. Kuiper predicted that particles of rings are composed of ice. Cassini probe (2004-2017) measured that particles mostly consists of water ice, 93% [2] and 90 to 95% [3].

There are several theories that have been developed to date to explain the evolution and stability of Saturn's rings postulate that the orbits of the rings are close to the equatorial plane of the planet and particles of rings are separated and do not collide each other, but none of approaches consistently explained these peculiarities [4].

Existing gravitational models of the Saturn's rings origin are based on the following assumption: a moon of the planet could have been disrupted by a passing celestial body; the rings could have been generated by the particles separated from moons of the outer planets by collision with comets or meteorites; the ring particles can be debris of a large comet tidally broken by the planet; the rings can be the relic of a protosatellite disk; the particles can be continuously forming as a result of volcanic activity on a moon of the Saturn. But none has provided a convincing explanation for rings observed phenomena entirely confirmed by Cassini probe [5].

## 2. The mechanism of magnetic anisotropic accretion

To determine correctly influence of the magnetic field on the diamagnetic particles of the protoplanetary cloud, we considered movement of these particles in the protoplanetary cloud with the simultaneous interaction of particles with gravitational and magnetic fields of Saturn.

This also leads us to an additional mechanism of magnetic anisotropic accretion due to the appearance of additional third force, which is the force of diamagnetic expulsion of a diamagnetic ice particle [6]. This force begins acting on the particles within protoplanetary cloud after emerging of the magnetic field of Saturn together with the gravitational force and centrifugal one.
After emerging of Saturn's magnetic field and appearance of the force of diamagnetic expulsion for the ice particles inside of protoplanetary cloud, all chaotic orbits of particles start to shift to the magnetic equator plane, because energy of particles have the minimum value on the magnetic equator. At the end of the movements, all the diamagnetic particles from the protoplanetary cloud come to the plane of the magnetic equator, which coincides with the geographical equator for Saturn. All particles are trapped in a three-dimensional magnetic well. Every particle on the magnetic equator comes to a stable position, where its horizontal and vertical shift is prevented. It means that the protoplanetary cloud is gradually collapsing into a disk of rings.

Interesting fact is taking into account the magnetic field of Saturn explains the transformation of a protoplanetary cloud into a disk of rings as well as it accounts for strong planar structure of rings located at the magnetic equator of Saturn which is almost coincide with geographical one. It also helps understand separation and do not sticking together of diamagnetic particles after formation of the disk of rings as well as to understand sharp edges of the rings.

Thus it becomes to be clear that Saturn could create rings itself from the diamagnetic ice particles of the protoplanetary cloud with the help of its own magnetic field due to the additional action of the third force of diamagnetic expulsion and due to the additional mechanism of magnetic anisotropic accretion [6].

## 3. Simulation of the motion of diamagnetic ice particles of the protoplanetary cloud under the common action of the gravitational and magnetic fields of Saturn

Our goal is to understand how protoplanetary cloud filled with diamagnetic ice particles can be transformed into a disk of rings (scenario on Fig. 1).We use theory of V. Safronov [7].

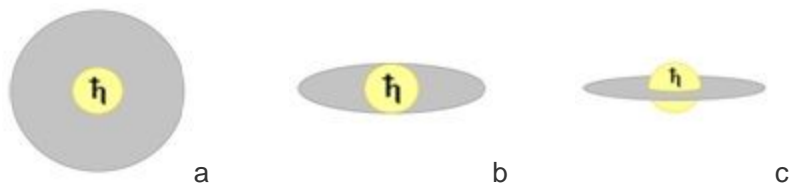

  a                          b                          c

Fig. 1. Transformation of Saturn's protoplanetary cloud into a disk of rings after common action of gravitational and magnetic fields of Saturn: from (a) >> (b) >> to (c)

Also necessary have in mind that for Saturn we have a spherically symmetric gravitational field and axisymmetric magnetic field. The mathematical solution of the problem is based on the fundamental theory [8]. Solution of the problem of the motion of diamagnetic particles after the emerging of the magnetic field of Saturn leads us to the equation for the azimuthal angle of the particles:

$$\ddot{\theta}_{S-p} + \dot{\theta}_{S-p} \cot\theta_{S-p} = \frac{GM_S}{r_{S-p}^3} \cot\theta_{S-p} + \frac{3C\mu_0^2 m_S m_p}{2\pi^2 r_{S-p}^8 M_p} \cot\theta_{S-p} \cos^2\theta_{S-p}$$

The first member on the right is responsible for gravity, the second - for the magnetic field. Here we see all orbits of ice particles at the end of their movement entering to magnetic equator plane. Solution $\Theta = \pi/2$ accounts for strong planar structure of Saturn's rings and their location in the magnetic equator plane.

And the equation for azimuthal velocity of particles in this case is:

$$\dot{\varphi}_{S-p} = \sqrt{GM_S/r_{S-p}^3 + 3C\mu_0^2 m_S m_p/(8\pi^2 r_{S-p}^8 M_p)}$$

Here we see the gravity force in the particle's orbit is counterbalanced by both the centrifugal force and the force of diamagnetic expulsion.

## 4. Stability, sharp edges of the rings and separation of particles

The entire model of Saturn's rings structure we considered as spatially separated structure of dense packaging particles in the Saturn's rings we presented as uniformly magnetized spheres in a disk-shape structure consisting of the same spheres. In result magnetization and magnetic moment of disk-like structure of rings is much higher than those of a single sphere due to the alignment of the multiple magnetic dipoles with the field. In addition, in a disk-shaped structure, the force of diamagnetic expulsion in the weak field region is stronger, and the magnetic well on the magnetic equator is deeper. These features provide strong stability of particles in the orbits and the entire ring system.

Magnetic well of the disc of rings with ice particles changes the pattern of magnetic field lines of Saturn, Fig. 2. The ice particles of rings are captured at the magnetic equator of Saturn in a magnetic well. It is very stable position for all particles: vertically it is due to minimum of the energy of particles at the equator, horizontally it is due to inhomogeneity of the magnetic field along radius.

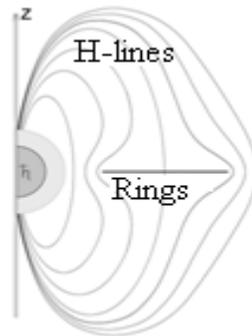

Fig. 2. Deformation of magnetic field lines of Saturn by the disc of rings

We suggest explanations of sharp edges of Saturn's rings and separation of their particles. Earlier explanation of these phenomena in the rings structure is based on the synchronization phenomenon due to which the epicycle rotational phases of particles in the ring, under certain conditions, become synchronized with the phase of external satellite [1]. In accordance data of Cassini probe existing explanation is valid only for irregularity of the edges of rings due to gravitational interaction of particles with satellite. At the same time sharp edges of rings and separation of particles contains even under influence of fluctuation of the gravitational field of Saturn and creation of wave of density and surface waves in the rings structure.

Diamagnetic ice particles of protoplanetary cloud collapse into the stable system of rings consisting of diamagnetic particles. Redistribution of particles inside the plane of rings is a result of their interactions between the areas of the magnetic flow gradient within the plane of the magnetic equator by the different force [9]. The forces are following: $F = -\mu dH/dz$, where $\mu$ – the magnetic moment of particle, $dH/dz$ – the gradient of the magnetic field along the z axis of the magnetic dipole. The force of the diamagnetic expulsion forms sharp edges of the ring: $F = -\mu dH/dy$, where $dH/dy$ – the gradient of the magnetic field along the radius of the ring. The accidental break in the ring will be stabilized by the force of the diamagnetic expulsion $F = -\mu dH/dx$, where $dH/dx$ - gradient of the magnetic field in the tangential direction.

The magnetic field in the plane of the rings disc is inhomogeneous. Magnetic field lines will strive to go through the region with the highest magnetic permeability and particles gather in the areas with low density of the magnetic field. Density gradient of magnetic field flow repel particles of each other, and also cleans the gaps within the rings system and forms a rigid thin structure of the separated rings. The density of the magnetic field flow inside each ring will be lower than within surrounding its space. The difference of density of the flow will cause directed inward magnetic pressure on the each ring, therefore, the rings have sharp edges (please, see Fig. 3)

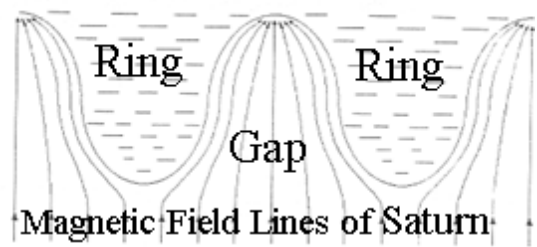

Fig. 3. Dense and rarefied areas of particles look like a system of rings

## 6. Notes about diamagnetism of particles in the Saturn's rings

If we compare the ratio of the thickness of rings to their diameter with a thickness of a sheet of paper to its length, the relative thickness of the disc rings will be a thousand times less. It is surprising fact that such a thin film of ice particles of huge diameter hangs in outer space. The question arises about the possible role of magnetic interaction in the origin of the rings, which have not yet been described. Therefore, we try to use diamagnetism of ice particles inside protoplanetary cloud to understand of observed features in the Saturn's rings.

Important fact is Cassini probe measured the ratio of deuterium and hydrogen isotopes for the ice of Saturn's rings is the same as for the Earth's ice [10] This fact indicates the similarity of ice in the rings and Earth's ice. We can choose ice XI [11], which is suitable for the environmental data of Saturn's rings. It has stable parameters below 73K and it is diamagnetic [12].

**Conclusion**

As a result, we have demonstrated the diamagnetism of the ice particles of a protoplanetary cloud can explain the sharp edges of the rings and separation of its ice particles.